\documentclass[12pt,letterpaper]{amsart}
\usepackage{amssymb,latexsym, amsmath, amsxtra}
\usepackage[dvips]{graphics}

\theoremstyle{plain}
\newtheorem{theorem}{Theorem}[section]

\newtheorem{lemma}[theorem]{Lemma}

\theoremstyle{definition}

\theoremstyle{remark}

\numberwithin{equation}{section}
\numberwithin{theorem}{section}

\newcommand{\Z}{\mathbb Z}

\def\({\left(}
\def\){\right)}

\def\P{F}

\begin{document}

\title[]{Crystallization of random trigonometric polynomials}%
\author{David W. Farmer \and Mark Yerrington}%
\address{}%
\email{farmer@aimath.org}%
\email{my001m@mail.rochester.edu}%

\thanks{Research of both authors supported by the
American Institute of Mathematics and
the NSF Focused Research Group grant DMS 0244660}

\subjclass{}%
\keywords{}%

\commby{AIM}
\begin{abstract}
We give a precise measure of the rate at which repeated differentiation 
of a  random trigonometric polynomial causes the roots of
the function to approach equal spacing.  This can be viewed as
a toy model of crystallization in one dimension.
In particular we determine the asymptotics of the distribution
of the roots around the crystalline configuration and find that
the distribution is not Gaussian.
\end{abstract}

\maketitle

\section{Introduction}

The critical points of an analytic function have a variety
of interesting physical interpretations.  For polynomials
we have the Gauss electrostatic model:  at each zero of the polynomial
$f(z)$
place identical point
charges obeying an inverse linear law.  Then the zeros of the derivative
$f'(z)$ are the points where the field vanishes.  To see when this works,
just write $f(z)$ in factored form and consider 
the logarithmic derivative $f'(z)/f(z)$.

\begin{figure}[htp]
\begin{center}
\scalebox{1.4}[1.4]{\includegraphics{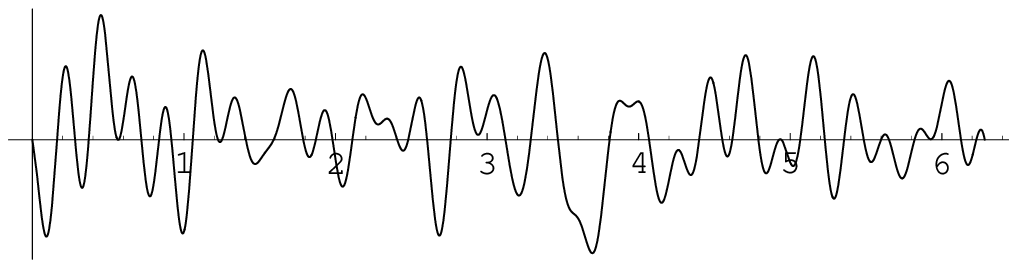}}
\vskip 0.15in
\scalebox{1.4}[1.4]{\includegraphics{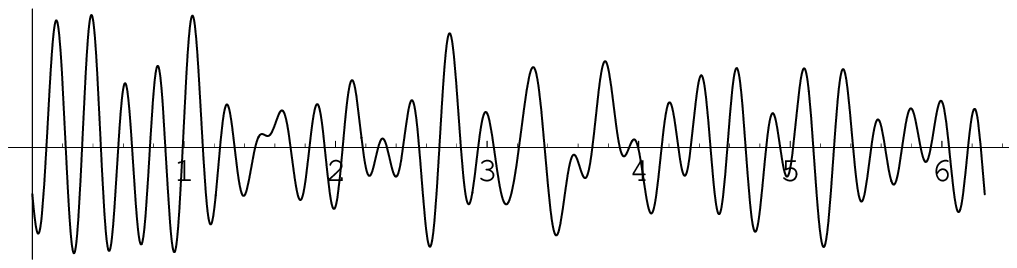}}
\vskip 0.15in
\scalebox{1.4}[1.4]{\includegraphics{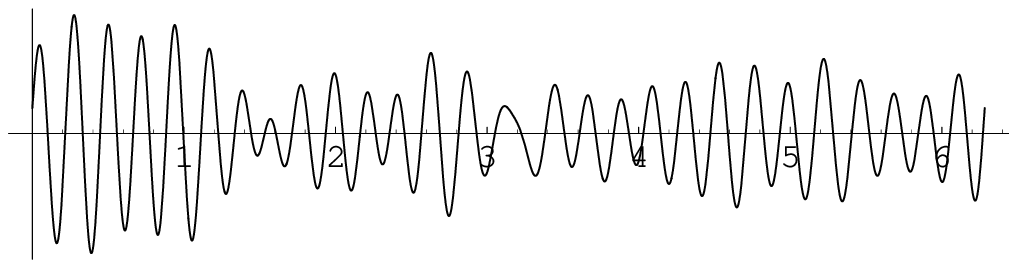}}
\vskip 0.15in
\scalebox{1.4}[1.4]{\includegraphics{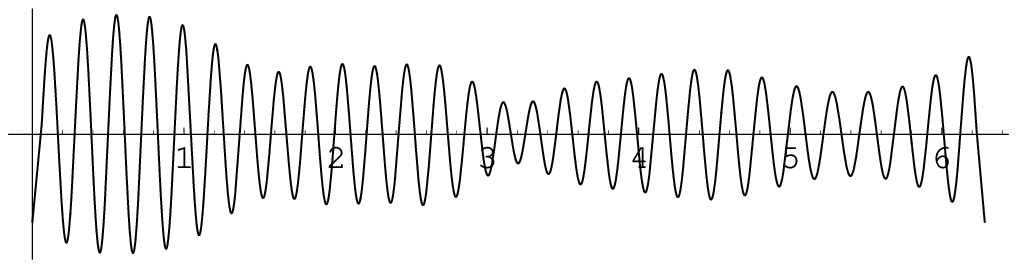}}
\caption{\sf
An example degree 30 trigonometric polynomial, along
with its 1st, 3rd, and 10th derivative.
} \label{fig:fplots}
\end{center}
\end{figure}

The Gauss model extends to entire functions of order~$1$, provided one
incorporates a background field coming from the exponential factors
of the Hadamard factorization of the function. For such functions
there is a general phenomenon that differentiation 
smooths out irregularities in the distribution of zeros.
See~\cite{FR} for details.  If the zeros are located in a strip around the
real axis and their initial distribution is not too irregular,
then repeated differentiation leads the zeros to approach equal
spacing.  This regular spacing is known as the
\emph{crystalline configuration}~\cite{BBL} and we view the process of
repeated differentiation as a toy model of crystallization in one dimension.
According to the Gauss model, regular spacing is the equilibrium position
and differentiation moves the function toward equilibrium.
See Figure~\ref{fig:fplots} for an illustration.

We now describe the functions we study and then discuss our results.

A random trigonometric polynomial of degree $N$ is a function of the form
\begin{equation}\label{eqn:rtp}
\P(x)=\sum_{n=0}^Na_{n}\cos(nx)+b_{n}\sin(nx)
\end{equation}
where $a_n$ and $b_n$ are random variables.  In this paper
we will assume that the $a_n$ and $b_n$ are independent real  Gaussian
distributed with mean $0$ and variance~$\sigma_n$, and usually
we further assume that the variances $\sigma_n$ are all equal.

Our concern is with the properties of the \emph{real} zeros
of~$\P(x)$.  Generally $\P(x)$ will not have all of its zeros
real, but the high derivatives of $\P(x)$ will have mostly real zeros,
and those zeros will be close to equally spaced.
This can be seen in Figure~\ref{fig:fplots}.
This is a general property~\cite{FR} of real entire
functions of order~$1$, but
in the case of trigonometric polynomials this is easy to see.
The $p$th derivative of~$\P(x)$ is
\[
\P^{(p)}(x)=\sum_{n=1}^N a_{n}n^p\cos(n x)+b_{n}n^p\sin(n x),
\]
where for simplicity we have assumed $p$ is a multiple of~$4$,
For large $p$ the terms $a_{N}N^p\cos(N x)+b_{N}N^p\sin(N x)$ dominate.
So the zeros of $\P^{(p)}(x)$ are close to the zeros of
$a_{N}\cos(N x)+b_{N}\sin(N x) = c_N \cos(N x + \phi_N)$ for some
real $c_N$ and $\phi_N$, and those zeros are real and equally spaced.

Another general property of repeated differentiation of 
real entire functions of order~$1$ is that
the discrepancy from equal spacing of zeros of the $p$th
derivative scales as~$O(1/p)$.  See Theorem~2.4.2 of~\cite{FR}.
In the case of random trigonometric
polynomials we are able to obtain more precise information.
We describe this in the next section.

\section{Statement of results}

In this section $\P(x)$ is a random trigonometric polynomial of
the form~\eqref{eqn:rtp} for which the $a_k$ and $b_k$ are
independent Gaussian distributed random variables with 
mean 0 and identical variance.
We wish to measure the rate at which the real zeros of 
the $p$th derivative $\P^{(p)}(x)$
approach equal spacing as~$p\to\infty$.  We consider the
pair correlation function of the zeros of  $\P^{(p)}(x)$,
defined as
\begin{equation}
R_{2,p}(\tau) =\langle \rho_p(x)\rho_p(x+\tau) \rangle ,
\end{equation}
where
\begin{equation}
\rho_p(x)=\sum_{x_k:\P^{(p)}(x_k)=0} \delta(x-x_k) 
=
\delta(\P^{(p)}(x)) |\P^{(p+1)}(x)| .
\end{equation}
Here $\delta$ is the Dirac $\delta$-function at~$0$, and
$\langle\cdot\rangle$ stands for expected value.
Thus, $\rho_p$ is the density function of real zeros of~$\P^{(p)}$,
and $R_{2,p}$ is the density function of differences of real zeros.

\begin{figure}[htp]
\begin{center}
\scalebox{0.70}[0.70]{\includegraphics{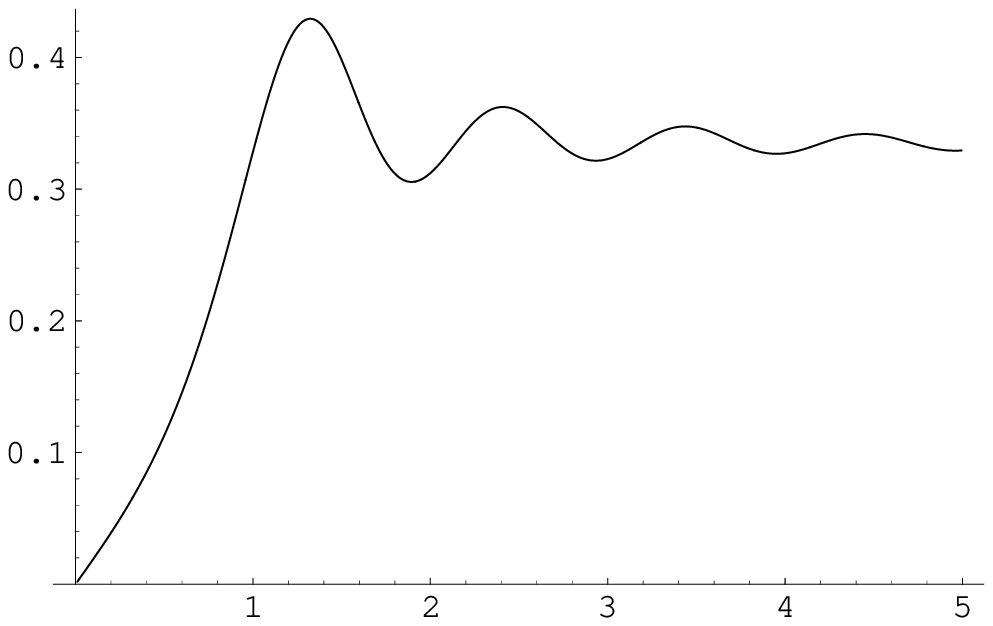}}
\hskip 0.3in
\scalebox{0.70}[0.70]{\includegraphics{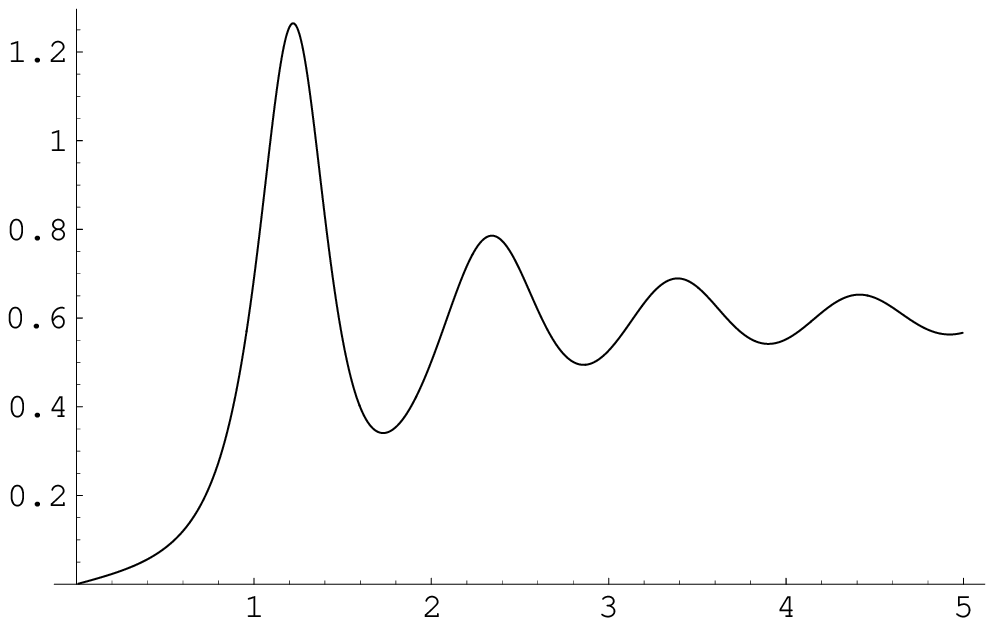}}
\vskip 0.15in
\hskip 0.1in
\scalebox{0.70}[0.70]{\includegraphics{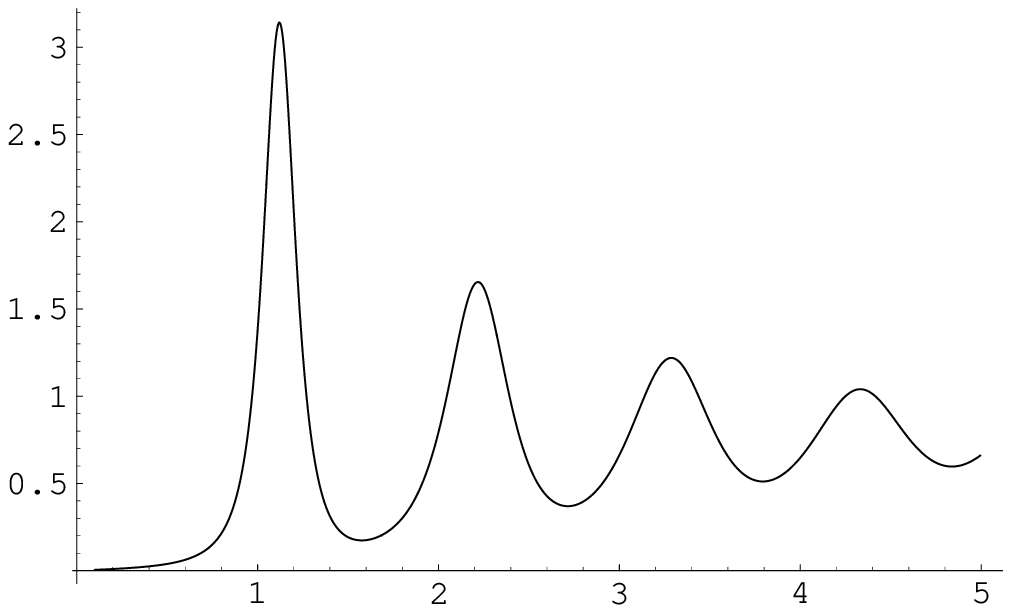}}
\hskip 0.35in
\scalebox{0.70}[0.70]{\includegraphics{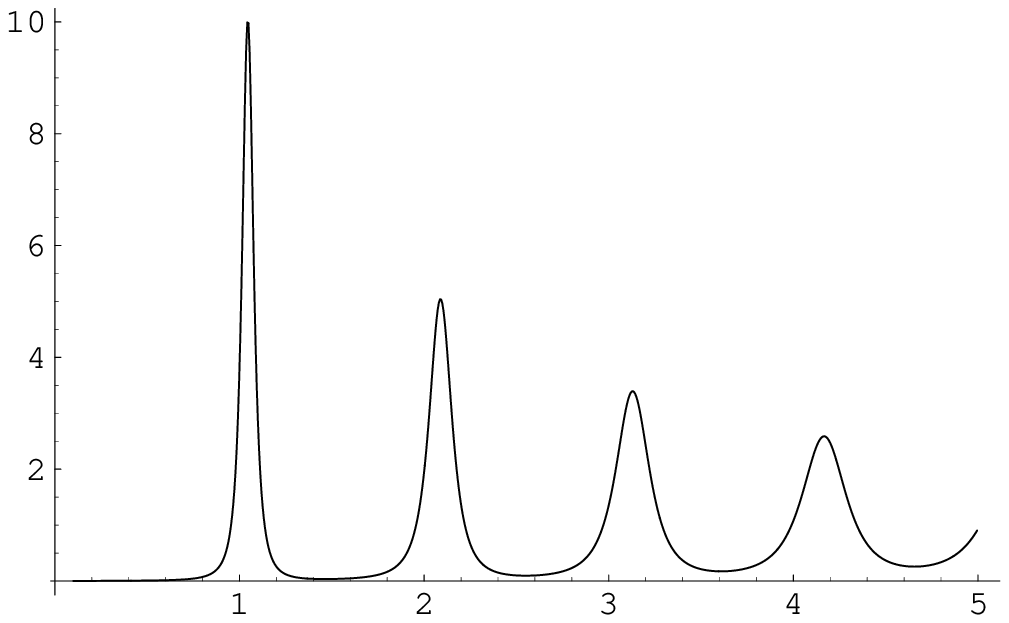}}
\caption{\sf
Plots of $R_{2,p}(x)$, the pair correlation funtion of the zeros
of the $p$th derivative $\P^{(p)}(x)$, for $p=0$,~$1$ (top row)
and $p=3$,~$10$ (bottom row).
} \label{fig:R01310}
\end{center}
\end{figure}

Since $R_{2,p}$ measures the differences between zeros, if the zeros
are almost regularly spaced then $R_{2,p}(x)$ will be large 
when $x$ is close to a multiple of the average zero spacing and it
will be small otherwise.  In other words, we expect that 
$R_{2,p}$ should approach a sum of $\delta$-functions at the
integers as $p\to\infty$.
Bogomolny, Bohigas, and Leb\oe uf~\cite{BBL} obtain a general expression
for the pair correlation function of the real zeros of a random trigonometric
polynomial.  From their results (which we describe 
in Section~\ref{sec:paircorrelation}) we obtain a formula for $R_{2,p}$,
which we plot for $p=0,1,3,10$ in Figure~\ref{fig:R01310}.
Note that the $p=0$ case is from~\cite{BBL}.

The plots in Figure~\ref{fig:R01310} use the following normalization.
We rescale the polynomial so that the average spacing between zeros
is~$1$, that is, we are actually considering the
function~$F^{(p)}(\pi x/N)$. As we describe in Section~\ref{sec:rtp},
as $N\to\infty$ the function $F^{(p)}$ has the expected fraction
\begin{equation}\label{eqn:fractionreal}
v_p =\sqrt{\frac{\mathstrut 2p+1}{\mathstrut 2p+3}} \sim 1-\frac{1}{2p}
\end{equation}
of real zeros.   
So $1/v_p$ is the average gap between consecutive real zeros, 
and that is the spacing between the peaks in the pair correlation functions
in Figure~\ref{fig:R01310}.   Note that $v_p$ is also the density of
real zeros.
So the pair correlation function $R_{2,p}(x)$ will equal $v_p^2$ on
average, which can also be seen in Figure~\ref{fig:R01310}.

We give an asymptotic formula
for the pair correlation function $R_{2,p}(x)$ as~$p\to\infty$.
As in the general case~\cite{FR} we find a $O(1/p)$ discrepancy from 
equal spacing, and furthermore
we find that the nearest-neighbor spacing has (appropriately rescaled)
distribution function
\begin{equation}\label{eqn:nn}
\frac{1}{(1+4 x^2)^{\frac{3}{2}}} ,
\end{equation}
centered at $1+\frac{1}{2p}$.
In particular, the discrepancy from equal spacing is not Gaussian.
The precise statement is
\begin{theorem}\label{thm:R2lim} As $p\to\infty$, the pair correlation function
 $R_{2,p}(x)$
of the real zeros of $\P^{(p)}$ approaches a sum of Dirac $\delta$-functions
at the nonzero integers.  The $\delta$-function near
the positive integer $n$ is given by
\begin{equation}
R_{2,p}\left(n\left(1+\frac{1}{2p}+ \frac{u}{p}\right)\right)
= \frac{p}{n} \frac{1}{(1+4 u^2)^{\frac32}} + O(1),
\end{equation}
as $p\to\infty$.
\end{theorem}

The proof is given in Section~\ref{ssec:proofofthm}
Note that the total area under $1/(1+4 u^2)^{\frac32}$ is~$1$, which shows
that the above Theorem does in fact identify the $\delta$-function near
the integer~$n$.

Since the zeros of $\P^{(p)}$ are close to equally spaced,
the peak of $R_{2,p}(x)$ near $x=1$ is almost completely due
to nearest-neighbor spacings.  Thus, we can read the nearest-neighbor
distribution from the pair-correlation function, as given in~\eqref{eqn:nn}.
In fact, we can also read off the next-nearest neighbor spacing 
(and all of the other neighbor spacings), as $p\to\infty$, from the
pair correlation function.  Up to rescaling, all of those distributions are
the same.  This shows that there are long-term correlations between
the zeros, otherwise, for example, the next-nearest neighbor 
distribution would be the convolution of the nearest neighbor 
distribution with itself. 

One motivation for this work is to understand, in general, the 
effect of differentiation on the statistics of zeros for functions
which have all their zeros on a line.
In particular, we would like to understand the effect
of differentiation on the repulsion between zeros.  For example,
if $f(x)$ has only real zeros and the zeros have the same statistics as the
eigenvalues of the classical random matrix $\beta$-ensemble,
then the zeros of $f(x)$ have repulsion of order $\beta$.
For the derivative~$f'(x)$,
it is reasonable to
conjecture that the zeros would have repulsion of order~$3\beta+1$, 
because pairs of close zeros of $f'(x)$ occur when $f(x)$ has three
closely spaced zeros.
This topic is of interest to number theory~\cite{FG}.
Unfortunately, the calculations in this paper do not shed light
on this phenomenon because the zeros
of random trigonometric polynomials are not in general all on the
real line.  We find that all derivatives of a random trigonometric
polynomial have \emph{linear} repulsion between zeros, and
show that this is due to the fact that each derivative moves new
zeros onto the real line, and those new zeros show linear repulsion
from each other.  See Section~\ref{ssec:linear}.


In the next section we discuss generalities about random trigonometric polynomials.
In Section~\ref{sec:paircorrelation} we discuss the asymptotics of the pair 
correlation of the zeros of $\P^{(p)}$, and in Section~\ref{ssec:linear}
we compute the repulsion between the zeros.

\section{Random trigonometric polynomials}\label{sec:rtp}

We assume that $\P(x)$ is a random trigonometric polynomial of
the form~\eqref{eqn:rtp}, where the $a_k$ and $b_k$ are
independent real normally distributed random variables with
mean 0 and identical  variance~$\sigma^2$.  

The expected fraction, $v_p$, of real zeros of $\P^{(p)}(x)$
is given in~\eqref{eqn:fractionreal}.  This result is
due to Dunnage~\cite{D}.  
It follows directly from  the Kac-Rice formula,
which we give in Lemma~\ref{lem:krf}.

Note that the original polynomial $\P(x)$ has on average
$1/\sqrt{3} \approx 57.7\% $ real zeros, the 4th 
derivative has more than 90\% of its zeros real, and one must
take 49 derivatives in order to expect 99\% real zeros.
Note that these are asymptotic results, and one can obtain 
exact formulas for any~$N$.  For example, the $N=30$, $p=10$
example at the bottom of Figure~\ref{fig:fplots} expects
to have 96.96\% real zeros, and this is approximately 1.4\%
larger than the asymptotic estimate~\eqref{eqn:fractionreal}.

We now derive~\eqref{eqn:fractionreal}.
 
\begin{lemma}{(The Kac-Rice Formula \cite{K,R})}\label{lem:krf}
Suppose $\P(x)$ is a random trigonometric polynomial with
independent real normally distributed coefficients having mean~0
but not necessarily equal variance, 
and let
\begin{align*}
A^{2}=\mathstrut&Var(\P(x))=\langle\P(x)^2\rangle\\
B^{2}=\mathstrut&Var(\P'(x))=\langle\P'(x)^2\rangle\\
C=\mathstrut&Cov(\P(x)\P'(x))=\langle\P(x)\P'(x)\rangle\\
\Delta^2=\mathstrut&A^2B^2-C^2.
\end{align*}
Then the expected number of real roots of $\P(x)$ in the interval
$(a,b)$ is
\[
\frac{1}{\pi}\int_a^b\frac{\Delta}{A^2}dx.
\]
\end{lemma}

In the case of $\P(x)$ of the form~\eqref{eqn:rtp} we have
\begin{align*}
A^{2}=\mathstrut&\sum_{n=0}^N \sigma_n^2 \\
B^{2}=\mathstrut&\sum_{n=0}^N n^2 \sigma_n^2 \\
C=\mathstrut&0.\\
\end{align*}
If all the coefficients of $\P(x)$ have equal 
variance~$\sigma^2$, then the $p$th derivative $\P^{(p)}(x)$
can be viewed as a random trigonometric polynomial of the 
form~\eqref{eqn:rtp} where the 
coefficients $a_n$ and $b_n$ have variance~$n^{2p} \sigma^2$.
So for the $p$th derivative we have
\begin{equation}
A = \sigma^2 \sum_{n=0}^N n^{2p} \sim \sigma^2 \frac{N^{2p+1}}{2p+1}
\end{equation}
and
\begin{equation}
B = \sigma^2 \sum_{n=0}^N n^{2p+2} \sim \sigma^2 \frac{N^{2p+3}}{2p+3},
\end{equation}
as $N\to\infty$.
From this, formula \eqref{eqn:fractionreal} for the fraction of
real roots $v_p$ follows immediately.

\section{Pair correlation of the real roots}\label{sec:paircorrelation}

We use a result of Bogomolny, Bohigas, and Leb\oe uf~\cite{BBL}
to compute the pair correlation function $R_{2,p}$ of the
real zeros of~$\P^{(p)}(x)$.

\begin{lemma}[Appendix B of \cite{BBL}]\label{l:corr}
Suppose 
\[
\P(x)=\sum_{n=0}^Na_{n}\cos(nx)+b_{n}\sin(nx)
\]
is a random trigonometric polynomial, where the 
$a_n$ and $b_n$ are independent real Gaussian distributed random
variables with mean 0 and variance~$\sigma_n$.
The expected value of the pair correlation function
$R_2(\tau)$ of the \emph{real} zeros of $\P(x)$ is
given by
\begin{equation}\label{eq:corr}
R_{2}(\tau)=\frac{1}{\pi^2 C^{\frac32}}\left(
B \arcsin(B/A)+\sqrt{A^2-B^2}\right),
\end{equation}
where 
\begin{equation}\label{eq:abz}
\begin{split}
A&=g_{2} C -g_{1}g_{4}^2\\
B&=g_{5} C-g_{3}g_{4}^2\\
C&=g_{1}^2-g_{3}^2,
\end{split}
\end{equation}
with
\begin{equation}\label{eq:gs}
\begin{split}
g_{1}&=\sum_{n=1}^N\sigma_n^2\\
g_{2}&=\sum_{n=1}^N n^2 \sigma_n^2 \\
g_{3}&=\sum_{n=1}^N \sigma_n^2 \cos(n\tau)\\
g_{4}&=\sum_{n=1}^N n \sigma_n^2 \sin(n\tau)\\
g_{5}&=\sum_{n=1}^N n^2 \sigma_n^2 \cos(n\tau).
\end{split}
\end{equation}
\end{lemma}

We apply the Lemma to $\P^{(p)}(x)$ 
with $\sigma_n=n^p$.
So we have 
\begin{equation}
g_1\sim \frac{N^{2p+1}}{2p+1} 
\ \ \ \ \ \ \ \ \ \ \ 
\text{and}
\ \ \ \ \ \ \ \ \ \ \ 
g_2\sim \frac{N^{2p+3}}{2p+3} ,
\end{equation}
as $N\to\infty$.
To determine asymptotics for $g_3$, $g_4$ and $g_5$, we
use the fact that for continuous functions $f$ 
\[
\int_0^1f(x)dx=
\lim_{N\rightarrow\infty}\frac{1}{N}\sum_{n=0}^Nf\left(\frac{n}{N}\right).
\]
We change variables $\tau=\pi x/N$, so in the variable $ x$
the mean spacing between zeros of $\P$ is unity.  We have
\begin{align}
g_{3} \sim &\mathstrut N^{2p+1}\int_0^1\cos\left(\pi x t\right)t^{2p}dt\cr
g_{4} \sim &\mathstrut N^{2p+2}\int_0^1\sin\left(\pi x t\right)t^{2p+1}dt\\
g_{5} \sim &\mathstrut N^{2p+3}\int_0^1\cos\left(\pi x t\right)t^{2p+2}dt.\nonumber
\end{align}

We let $R_{2,p}( x)$ denote the pair correlation function of the
real zeros of the $p$th derivative $\P^{(p)}(x)$, where 
$\P(x)$ is given by \eqref{eqn:rtp}  with all $a_j$ and $b_j$ independent identical
Gaussian, and we normalize by dividing by the square of the overall
zero density~$N^2/\pi^2$.  As $N\to\infty$ we have
\begin{equation}\label{eq:corrp}
R_{2,p}( x) \sim \frac{1}{C_p^{\frac32}}\left(
B_p \arcsin(B_p/A_p)+\sqrt{A_p^2-B_p^2}\right),
\end{equation}
where
\begin{equation}\label{eq:abcp}
\begin{split}
A_p&=g_{2,p} C_p -g_{1,p}g_{4,p}^2\\
B_p&=g_{5,p} C_p-g_{3,p}g_{4,p}^2\\
C_p&=g_{1,p}^2-g_{3,p}^2,
\end{split}
\end{equation}
with
\begin{equation}\label{eq:gsp}
\begin{split}
g_{1,p}&=\frac{1}{2p+1}\cr
g_{2,p}&=\frac{1}{2p+3}\cr
g_{3,p}&=\int_0^1\cos\left(\pi x t\right)t^{2p}dt\\
g_{4,p}&=\int_0^1\sin\left(\pi x t\right)t^{2p+1}dt\cr
g_{5,p}&=\int_0^1\cos\left(\pi x t\right)t^{2p+2}dt.\cr
\end{split}
\end{equation}

Plots of $R_{2,p}$ for $p=0,1,3,10$ are given in Figure~\ref{fig:R01310}.

\subsection{Large $p$ asymptotics of the pair correlation}\label{ssec:proofofthm}

We determine the rate at which 
 $R_{2,p}( x)$, appropriately rescaled, approaches 
a sum of $\delta$-functions at the integers.

We first find asymptotic formulas for $A_p$, $B_p$, and $C_p$.
Using a geometric series expansion for $g_{1,p}$ and $g_{2,p}$,
and using integration by parts followed by a geometric series
expansion for  $g_{3,p}$, $g_{4,p}$, and $g_{5,p}$, we find
(with the help of a computer algebra package), that
\begin{align}
A_p=\mathstrut&
\frac{1}{64}\left(-2 \pi^2 x^2-2 \sin (2 \pi x) \pi x+\left(4 \pi^2 x^2-1\right) \cos (2 \pi x)+1\right)
{p^{-5}}
+O(p^{-6})
\cr
B_p=\mathstrut&
\frac{1}{128}\left(\cos (\pi x)+\left(4 \pi^2 x^2-1\right) \cos (3 \pi x)-8 \pi x \sin (\pi x)\right)
{p^{-5}} 
+O(p^{-6}) 
\cr
C_p=\mathstrut&\frac{1}{4}\sin ^2(\pi x)\,{ p^{-2}}
-
\frac{1}{4}\left( (\pi x \cos (\pi x)+\sin (\pi x)) \sin (\pi x)\right){p^{-3}} \cr
&+
\frac{1}{32}\left( \pi^2 x^2+8 \sin (2 \pi x) \pi x+3 \left(\pi^2 x^2-1\right) \cos
   (2 \pi x)+3\right) {p^{-4}}
+O(p^{-5})\cr
\end{align}

We see that $A_p$ and $B_p$ are generally of size $p^{-5}$,
while $C_p^{3/2}$ is generally of size~$p^{-3}$.  But for $x\in \Z$ 
we see that $C_p^{3/2}$ is of size~$p^{-6}$.
Thus, as $p\to\infty$, if $x\in \Z$ then $R_p(x)\to\infty$,
otherwise $R_p(x)\to 0$. This is exactly what one should expect
because
$R_p$ is approaching a sum of $\delta$-functions at the integers.
We now express this more precisely.

The minima of $C_p$ are not exactly at the integers, but they
are shifted over to approximately $n(1+\frac{1}{2p})$ for 
$n\in\Z$.  This is also what one would expect because the mean
spacing of the \emph{real} zeros of $\P^{(p)}$ is 
approximately $1/v_p \sim 1+\frac{1}{2p}$.  We will verify this directly 
from the 
above formulas.  Differentiating $C_p$ with respect to $x$ we have
\begin{equation}
C_p'(x)=
-\frac{1}{16}\left(\pi  \left((4 p-11) \pi  \cos (2 \pi  x) x-\pi  x+\left(-4 p^2+6 p+3 \pi ^2 x^2-7\right) \sin (2 \pi 
   x)\right)\right)p^{-4}. 
\end{equation}
We are thinking of $x$ as fixed and $p$ large, so the minimum of $C_p$
is close to the solution to
\begin{equation}
4 \pi p x \cos(2\pi x) - 4 p^2 \sin(2 \pi x)=0,
\end{equation}
which is equivalent to 
\begin{equation}
\tan(2\pi x)=\frac{\pi x}{p}.
\end{equation}
The solutions with $x$ near an integer correspond to the minima of~$C_p$,
so writing $x=n+\xi$ and $\tan(2\pi x)\sim 2\pi\xi$, we find
\begin{equation}
\xi\sim \frac{n}{2p} ,
\ \ \ \ \ \ \ \ \
\text{therefore}
\ \ \ \ \ \ \ \ \
x\sim n\(1+\frac{1}{2 p}\),
\end{equation}
as expected.

To show the shape of the $\delta$-functions, we expand near the
minima of $C_p$.  For positive integers $n$ we 
find that 
\begin{equation}
C_p\(n\(1+\frac{1}{2 p}+\frac{u}{p}\)\) = 
\frac{\pi^2}{16} \left(1+4 u^2\right) n^2 p^{-4} + O(p^{-5}).
\end{equation}
Also,
\begin{align}
A_p\(n\(1+\frac{1}{2 p}+\frac{u}{p}\)\) =\mathstrut&
\frac{\pi^2}{32} n^2 p^{-5} + O(p^{-7}) 
\cr
B_p\(n\(1+\frac{1}{2 p}+\frac{u}{p}\)\) =\mathstrut&
\cos(\pi n) \frac{\pi^2}{32} n^2 p^{-5} + O(p^{-7}).
\end{align}
Thus, $\arcsin(B_p/A_p) =(-1)^n \pi/2 + O(p^{-2})$, so we 
obtain Theorem~\ref{thm:R2lim}.

\subsection{Repulsion between zeros}\label{ssec:linear}

In \cite{BBL} it was found the the real zeros of $\P(x)$
have linear repulsion.  We find that the real zeros of
$\P^{(p)}(x)$ also show linear repulsion.  

\begin{figure}[htp]
\begin{center} \scalebox{0.70}[0.70]{\includegraphics{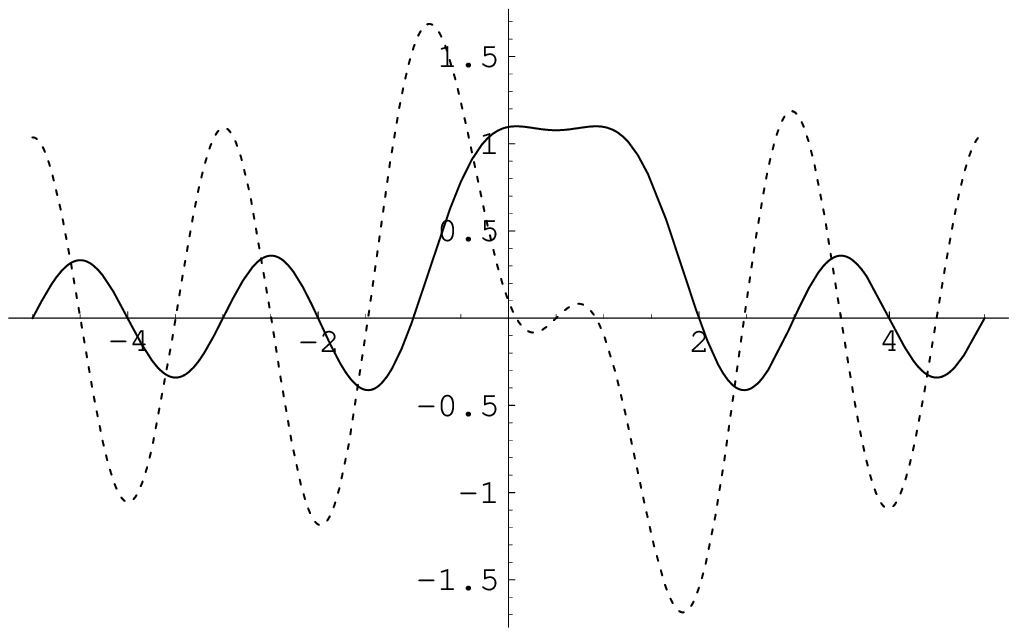}}
\hskip 0.3in
\scalebox{0.70}[0.70]{\includegraphics{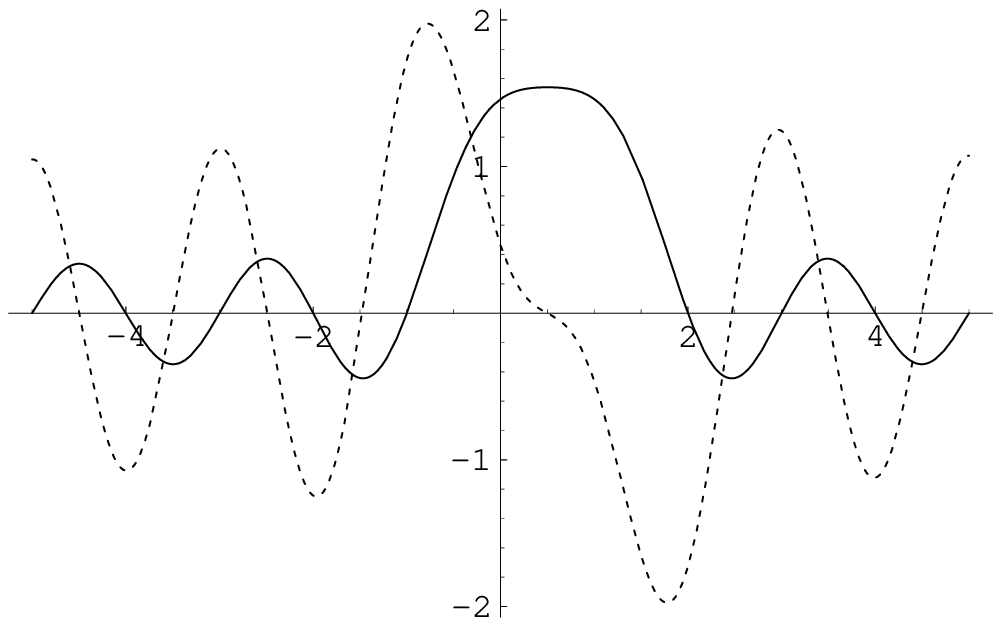}}
\caption{\sf Functions with zeros at $\frac12\pm i a$ and at the 
integers except for~$0,1$.  On the left $a=0.92$, and on the
right~$a=1.1$.  Dotted curve shows the derivative.
} \label{fig:linearrepulsion}
\end{center}
\end{figure}

Computing the small $x$ asymptotics of~\eqref{eq:gsp} we find that
\begin{align}
R_{2,p}(x)=\mathstrut&
\frac{\pi^2 \sqrt{\mathstrut 4 p^2+8 p+3} }{2 (2 p+3)^2 (2 p+5)} x
+
\frac{\pi^2 \left(4 p^2+8 p+3\right)^{3/2} }{(2 p+1) (2 p+3)^2 \sqrt{(2 p+5)^3 (2 p+7)}}x^2
+ O(x^3)\cr
\sim\mathstrut & \frac{\pi^2}{8p^2} x,
\end{align}
as $p\to\infty$.
So for every derivative we have \emph{linear} repulsion between zeros.
This appears to contradict the expectation that differentiation increases the
repulsion between zeros.  However, there is a simple explanation.
Each derivative causes more zeros to fall onto the real line.  
These ``new'' zeros can be closely spaced, as illustrated in 
Figure~\ref{fig:linearrepulsion}. 

In Figure~\ref{fig:linearrepulsion}, all of the zeros are spaced one unit
apart, except for two zeros which have been moved to have imaginary
part~$\pm a$ and real part the midpoint of the resulting gap.
In the left plot we have $a=0.92$ and in the right plot~$a=1.1$.
Moving the complex zeros to have imaginary part
\begin{equation}
\pm \frac{2}{\pi^2-8} \approx \pm 1.06975
\end{equation}
gives a triple zero of the derivative, so values slightly less
than this give closely spaced zeros.

Since
the fraction of new real zeros for the $p$th derivative is
\begin{equation}
v_{p}-v_{p-1}=\sqrt{\frac{2 p+1}{2 p+3}}-\sqrt{\frac{2 p-1}{2 p+1}}
\sim \frac{1}{2p^2} ,
\end{equation} 
which is of the same magnitude as the linear repulsion, we see that
the new zeros lead to repulsion of magnitude~$x/p^2$.
It seems reasonable to believe that
this accounts for all of the linear repulsion, but we have not been
able to verify this.

\bibliographystyle{amsplain}

\end{document}